*Review Article*

# Large Radio Telescopes for Anomalous Microwave Emission Observations

**E. S. Battistelli,[1] E. Carretti,[2] P. de Bernardis,[1] and S. Masi[1]**

[1] *Physics Department, Sapienza University of Rome, Piazzale Aldo Moro 5, 00185 Rome, Italy*
[2] *CSIRO, Astronomy and Space Science, P.O. Box 276, Parkes, NSW 2870, Australia*

Correspondence should be addressed to E. S. Battistelli, elia.battistelli@roma1.infn.it

We discuss in this paper the problem of the Anomalous Microwave Emission (AME) in the light of ongoing or future observations to be performed with the largest fully steerable radio telescope in the world. High angular resolution observations of the AME will enable astronomers to drastically improve the knowledge of the AME mechanisms as well as the interplay between the different constituents of the interstellar medium in our galaxy. Extragalactic observations of the AME have started as well, and high resolution is even more important in this kind of observations. When cross-correlating with IR-dust emission, high angular resolution is also of fundamental importance in order to obtain unbiased results. The choice of the observational frequency is also of key importance in continuum observation. We calculate a merit function that accounts for the signal-to-noise ratio (SNR) in AME observation given the current state-of-the-art knowledge and technology. We also include in our merit functions the frequency dependence in the case of multifrequency observations. We briefly mention and compare the performance of four of the largest radiotelescopes in the world and hope the observational programs in each of them will be as intense as possible.

## 1. Introduction

The interest of the scientific community in the Anomalous Microwave Emission (AME) is growing. This emission was first detected by the Cosmic Background Explorer (COBE) as diffuse emission excess at 30 GHz [1], and significant effort has been spent to understand its nature and properties since then. What is puzzling about it is that it is correlated with the brightness of thermal dust but peaks at 20–30 GHz where the thermal dust contribution is negligible and the emission budget is supposed to be dominated by free-free and synchrotron. It is thus a new type of emission, whose nature is still unknown. Its investigation is thus crucial for understanding the composition of the interstellar medium (ISM).

Statistical evidence of this emission has been found in observations of the diffuse Galactic background by several authors (see, e.g., [2–5]), while AME has been detected directly in individual regions in a limited number of cases (see Table 1), which still makes to derive its properties hard.

The AME regions typically emit 10%–30% more than expected from synchrotron and free-free, in the range 8–30 GHz (e.g., [6, 7], see Figure 1). Most of the published results relate to galactic objects, with some exceptions. The galactic sources are typically several arc minutes wide and are associated to HII regions, supernova remnants (SNRs) or dark clouds. Among them, only in one case, a significant polarization degree has been detected [8], while upper limits have been set for other cases [9–11]. Several models have been proposed to explain AME, the most popular of which are by Draine and Lazarian.

(i) Spinning dust (electric dipole): Draine and Lazarian [12, 13] (hereafter DL98a and DL98b) have proposed that the AME is due to electric dipole emission from rapidly rotating small dust grains in the ISM (i.e., spinning dust). They find that the emission spectrum produced by such grains fits well the observed signal and could be responsible for the AME. Lazarian and Draine [14] also found that paramagnetic relaxation resonance may be efficient at producing an alignment of grains rotating faster than 1 GHz. This may result in an observable polarization degree up to 5% at 10 GHz, decreasing at higher frequencies. Spinning dust models have then been refined since then [15–18].

Table 1: Individual galactic and extragalactic regions with reported detections of AME. The frequency and the resolution refer to the observations discovering the AME even when the same regions have been observed by other experiments at different frequency and angular resolution. Planck detections relate to the early release. Several other regions have already been announced in the intermediate release.

| Region | RA (J2000) | DEC (J2000) | Frequency (GHz) | Resolution | Experiment | Reference |
|---|---|---|---|---|---|---|
| NCP loop | | +88° | 14.5, 32 | 7'.35 | OVRO | Leitch et al. [38] |
| LDN1622 | 05 h 54 m 23 s | +01°46'54'' | 5, 8–10 | 6' | GBT | Finkbeiner et al. [34] |
| Perseus | 03 h 41 m 36 s | +31°48' | 11–17 | 0°8 × 1°1 | COSMOSOMAS | Watson et al. [6] |
| 3C396 | 19 h 04 m 04 s | +05°27'12'' | 33 | 9'.1 × 7'.7 | VSA | Scaife et al. [39] |
| RCW49 | 10 h 24 m 20 s | −57°44'57'' | 31 | 6' | CBI | Dickinson et al. [10] |
| LDN1688 | 16 h 25 m 57 s | −24°20'50'' | 31 | 6' | CBI | Casassus et al. [11] |
| RCW175 | 18 h 46 m 40 s | −03°46'00'' | 31 | 4' | CBI | Dickinson et al. [7] |
| LDN1111 | 21 h 40 m 30 s | +57°48'00'' | 14–18 | 2'.4 × 2'.1 | AMI | Ami Consortium [40] |
| LDN675 | 19 h 23 m 53 s | +11°07'39'' | 14–18 | 2'.4 × 2'.1 | AMI | Scaife et al. [41] |
| LDN1621 | 05 h 55 m 22 s | +02°11'33'' | 31 | 6'.1 × 4'.8 | CBI | Dickinson et al. [42] |
| LMC/SMC | 00 h 52 m/05 h 23 m | −72°48'/−69°45' | 23–94 | 1° | WMAP | Bot et al. [26] |
| NGC 6946 | 20 h 34 m 52 s | +60°09'14'' | 26–40 | 25'' | GBT | Murphy et al. [25] |
| rho Oph | 16 h 37 m 9 s | −10°34'01'' | 31 | 8' | CBI | Vidal et al. [22] |
| LDN1780 | 15 h 40 m 30 s | −07°14'18'' | 31 | 8' | CBI | Vidal et al. [22] |
| Pleiades | 03 h 47 m | +24°07' | 11–17, 23–94 | 1° | COSMOS./WMAP | Génova-Santos et al. [43] |
| AME-G173.6 + 2.8 | 05 h 41 m | +35°51' | 28–857 | 30' | Planck | Planck Collaboration [27] |
| AME-G107.1 + 5.2 | 22 h 22 m | +63°23' | 28–857 | 30' | Planck | Planck Collaboration [27] |

(ii) Dust grain magnetic dipole: Draine and Lazarian [19] show that magnetic dipole emission from dust grains made of strongly magnetic materials can have a spectrum accounting for the observed AME. For this model, ferromagnetic relaxation may efficiently align dust grains and produce strongly frequency and shape-dependent polarized emission, that could be as high as 30% at 10 GHz. This is further studied by Draine and Hensley [20].

Polarization level of the AME is a key point to select among models. Stokes I multiband observations in the range 8–30 GHz are instead essential to separate the AME from the other components and to confirm the rising spectrum of the AME. High angular resolution measurements, both in intensity and in polarization, are essential to understand the physics beyond the AME and to limit confusion arisen by the difficulty to disentangle different component emissions within the same region in poor angular resolution observations.

Recent models seem to predict that the AME is dominated by electric dipole emission from the smallest grains, possibly polycyclic aromatic hydrocarbons (PAHs) [21]. However, currently no observation has had sufficient sensitivity, resolution, and frequency coverage to disentangle the candidate mechanisms. This calls for further investigation of the properties of known anomalously emitting regions and to search for new regions with higher sensitivity, frequency coverage, and polarization measurements, to further increase our understanding of the physical process producing the AME. High angular resolution (arc minute level) observations are thus crucial to disentangle different contributions within the same region. These are starting to reveal surprising effects as the vanishing of the dust-to-radio correlation when we go to fine angular scales [22, 23]. This effect is explained by Ysard et al. [24] who explain how, in dense molecular clouds, the AME is stronger towards the center of the cloud due to the higher gas density, whereas the intense surrounding radiation field boosts the dust emission in the outskirts of the cloud, where the AME is lower due to the decrease on the gas density. This makes 50–100 m class telescopes ideal instruments for such observations.

## 2. Localized AME Measurements

High angular resolution measurements of anomalous emitters are important to distinguish different regions within HII regions or dark clouds, in order to separate the various components and understand the physics behind these mechanism isolating single emitters. In fact, the wide majority of the detected sources are within the galactic plane. In this case, arc minute level angular resolution would be efficient at disentangling the AME in the presence of galactic emission arising from magnetic fields where the different amount of dust, free electrons, and distributed magnetic field may act at mimicking rising spectra consistent with AME. Since AME is identified by means of SED measurements, high angular resolution measurements are essential for selecting, at the different frequencies, the same physical volume of emitter in the ISM. The same applies to the comparison of radio and IR-dust emission, essential to monitor small dust grains. To match the resolution of IR surveys, the largest radio telescopes must be used. When cross-correlating high- and low- frequency (and resolution) maps, one has to degrade the highly resolved map to the worst ones loosing information about the morphology of the emitting regions. It is useful and efficient to perform pixel-by-pixel correlation in order to extract emissivity information (see, e.g., [22]). In Figure 2, we show the effect of a loss of resolution in the MIPS 24-micron SPITZER maps of the Perseus complex when degrading it at 10 arc min resolution. Even more important is the angular resolution for extragalactic observations. Mapping the high star formation regions of near-by galaxies may improve our knowledge of the mechanisms as well as the comparison between ours and other galaxies. Individual regions observed so far to be anomalous emitters include HII regions, dark clouds, supernova remnants and high star formation rate regions in nearby galaxies. With the exception of NGC 6946, none of the measured regions have been observed, in the microwave range, at angular resolution beyond a few arc minutes. We report in Table 1 a list of published sources.

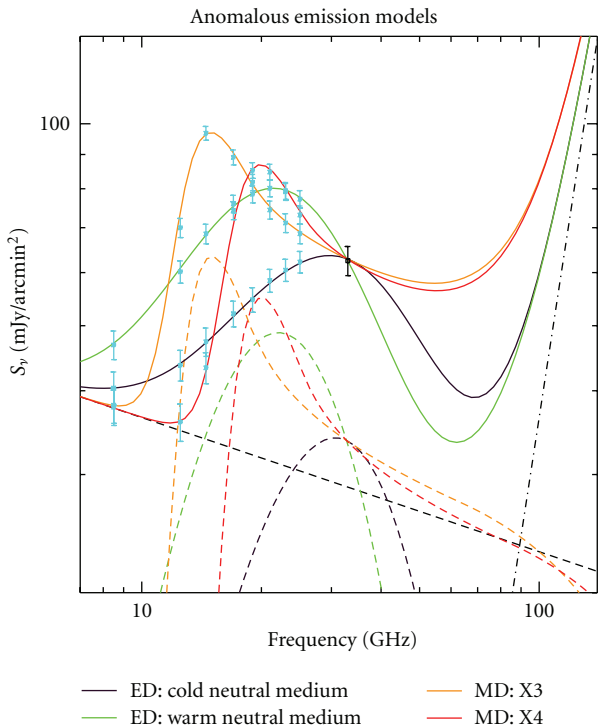

Figure 1: Hypothetical anomalously emitting region. The 33 GHz signal shown in the figure is inconsistent with the combination of free-free emission (dashed line) plus thermal dust emission (dot-dashed line), so an additional contribution is required to reach the measured flux. This is provided by different models of AME for electric dipole (ED, [13]) and magnetic dipole (MD, [19]). All plotted models are consistent at 33 GHz but are different at lower frequencies. Microwave measurements with high SNR will be able to select the correct model. The shown error bars correspond to the expected nominal sensitivity of the Parkes receivers.

All the reported cases are galactic objects with the noticeable exception of the nearby galaxy NGC 6946, which shows evidence of AME from an outer-disk star-forming region [25], and the tentative detection towards the Magellanic Clouds [26, 27]. Among them, only Perseus has been detected in polarization [8], while upper limits have been set for LDN1622, RCW49, and LDN1688 [9–11] as well as for the Perseus molecular cloud [28].

The Perseus molecular cloud is the most studied region: after the first detection by the COSMOSOMAS experiment [6], it has been observed by other experiments like VSA [29, 30] and Planck Collaboration [27]. Another very well-studied region is RCW175 [31] which seems to be a source of important information not only for its integrated spectral energy density (SED), but also of its complex morphology [32]. Angular resolution down to the 1 arc min level may be enough to separate the different contributions within the same region.

Measurements at frequencies sensitive to AME (10–40 GHz) in localized regions have to be compared to lower frequency measurements, typically up to 10 GHz in order to monitor the presence of synchrotron and free-free and to assess the excess of emission. Comparing total power continuum observations, often arising from different experiments, is both crucial and difficult. Different calibration schemes, observational strategies, map-making algorithms, and background removal techniques may result in biased results when comparing between the free-free and synchrotron signals to the AME. Multifrequency measurements are thus advisable and when performed with the same receiver would then be useful in order to control common systematic effects. In particular the choice of the observational frequency is of fundamental importance and has to be chosen trading off incoming power, detector sensitivity, observational contamination, and, when multifrequency measurements are undertaken, the steepness of the frequency spectrum together with the capability of disentangling power law decreasing spectra from the typical rising spectrum of AME.

In Figure 3(a), we plot the SNR in a typical ground-based observation build accounting for the atmosphere and receiver contribution to the system temperature (as a function of the frequency) as well as the signal intensity and spectrum. This can be considered a merit function for single frequency observation. For multifrequency measurements (Figure 3(b)), we built a similar merit function with the additional ingredient of aiming at disentangling AME and power-law emission like free-free or synchrotron. The combination of the two (Figure 3(c)) accounts for both the AME power as well as its spectrum.

## 3. Radio Telescopes for AME Observations

Continuum broadband measurements at radiofrequency (RF) with total power receivers are affected by systematic effects such as $1/f$ noise and gain fluctuations, stripes in maps, spikes in data, variation in the transmission, and emission of the atmosphere (all resulting in striping of the maps) [33]. Ambient environment condition changes as well as instrumental instability obviously influence the measurements. The final sensitivity clearly depends on the receiver sensitivity as well as its stability and the astronomical confusion.

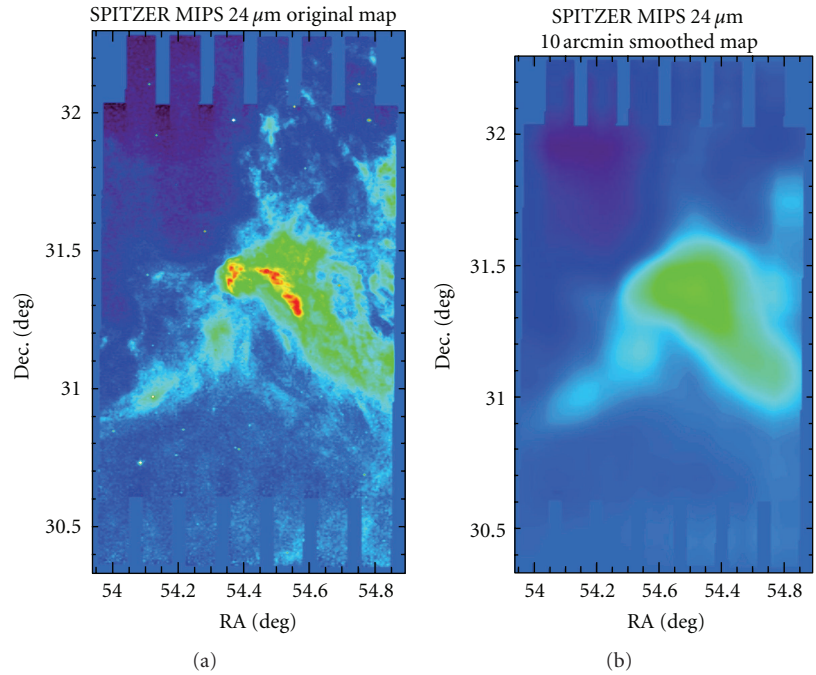

Figure 2: We report in this figure a 24 $\mu$m-MIPS SPITZER map of the Perseus molecular complex. In (a), the structures detected by the high angular resolution of SPITZER are evident. In (b), we see a degraded angular resolution map to 10 arc min in order to match current microwave resolution of typical AME regions. It is evident that the lower angular resolution introduces a shift in the emission centroid for nonspherical emitters. This results in a difficult interpretation in the comparison between microwave and IR maps detected at such different angular resolution.

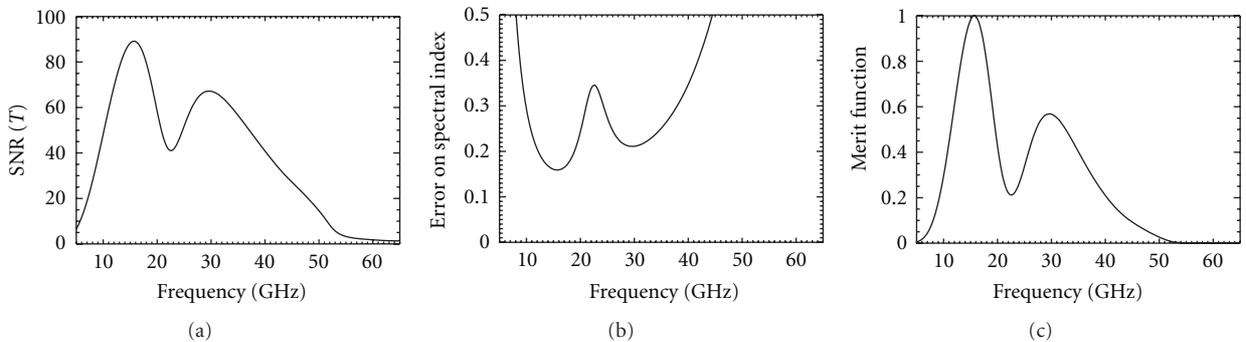

Figure 3: (a) SNR in ground-based observation of the SED of an AME cloud. The calculation includes the contributions of typical atmospheric and system noise and a typical spectrum of a warm ionized medium region. The dips in the SNR are due to atmospheric water (around 22 GHz) and oxygen (near 60 GHz). High SNR can be achieved with a long integration time and a large (50 m class) telescope. (b) Error in the estimate of the spectral index of the AME SED above, assuming that the same receiver is used at two frequencies 10% apart. This measurement is important if one wants to discriminate between AME and other contaminants (synchrotron or free-free or thermal dust). (c) Merit function computed as the ratio between the SNR and the error in the determination of the spectral index, normalized to the maximum. This indicates which is the optimal frequency to measure both the brightness of the AME and its spectral index.

Specific observational techniques are needed to mitigate the effect and include position switching (typically observing the sky a few beams apart), frequency switching (typically for spectral line observations, where the interested line is compared with the continuum baseline), Dicke switching (comparing the observed sky with an internal reference), IF switching (comparing 2 IFs), polarization switching (comparing different polarizations), cross-link scanning strategy (in which each sky region is observed at least twice with almost orthogonal scan directions), diodes illumination (used as reference signal to overcome drifts), detailed noise filtering, and 2/3 fields beam switching observations. Low latitude observatories may also take advantage, when performing maps, of the constant elevation scan strategy.

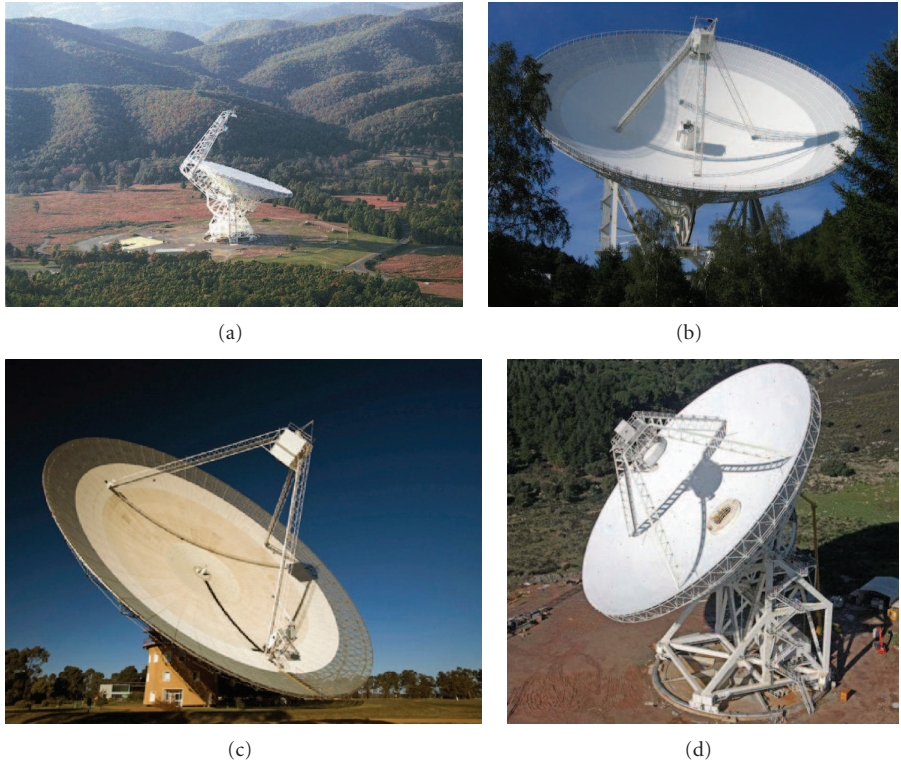

Figure 4: Radio telescopes appropriate for AME observations. From top left, clockwise, we see the Green Bank Telescope (https://science.nrao.edu/facilities/gbt/), Effelsberg (http://www.mpifr-bonn.mpg.de/8964/effelsberg/), Sardinia Radio Telescope (http://www.srt.inaf.it/), and Parkes (http://www.parkes.atnf.csiro.au/).

Table 2: Characteristics of the 3 radio telescopes able to perform continuum observations at 13 GHz and 23 GHz. Sensitivities are taken from the online sensitivity calculators assuming the same observation conditions.

|  |  | Parkes | Green Bank Telescope | Sardinia Radio Telescope |
|---|---|---|---|---|
|  | Frequency | 0.7 GHz–26 GHz | 0.1 GHz–116 GHz | 0.3 GHz–100 GHz |
| 13 GHz | Sensitivity | 3.7 mJy$\sqrt{s}$/beam | 0.8 mJy$\sqrt{s}$/beam | 0.6 mJy$\sqrt{s}$/beam |
|  | $T_{sys}$ | 150 K | 28 K | 29 K |
|  | Polarization capability | Yes | Yes | Yes |
|  | Angular resolution | 1.9 arc min | 0.95 arc min | 1.5 arc min |
| 23 GHz | Sensitivity | 3.2 mJy$\sqrt{s}$/beam | 1.7 mJy$\sqrt{s}$/beam | 2.0 mJy$\sqrt{s}$/beam |
|  | $T_{sys}$ | 95 K | 51 K | 81 K |
|  | Polarization capability | Yes | Yes | Yes |
|  | Angular resolution | 1.1 arc min | 0.54 arc min | 0.9 arc min |

The atmospheric emission tends, in constant elevation scan, to be dominated by the atmosphere small fluctuations rather than by large air-mass dependence. In order to remove stripes in the map, one can then use a cross-link strategy observing the same region when it rises and when it sets.

Among the instruments that we believe could drastically push ahead the knowledge of the AME mechanisms, we mention the 100 m Green Bank Telescope (http://science.nrao.edu/facilities/gbt/) in US, the German 100 m Effelberg telescope (http://www.mpifr-bonn.mpg.de/8964/effelsberg/), the Australian 64 m Parkes telescope (http://www.parkes.atnf.csiro.au/), and the new upcoming Sardinia Radio Telescope (http://www.srt.inaf.it/) in Italy. The telescopes are shown in Figure 4 and some general characteristics are reported in Table 2.

*3.1. Green Bank Telescope.* The Green Bank Telescope (GBT) is the largest fully steerable single-dish telescope in the world. Its off-axis design makes the GBT aperture to incoming radiation totally unblocked although increases the design complexity and the polarization interpretation due to the lack of circular symmetry. The active surface makes the GBT a suitable instrument up to 50 GHz as well as in the 80–100 GHz bands. It is located in a radio quite zone which makes it particularly free from interference. At the frequencies of interest for AME measurements, the GBT is equipped with X, K, Ku, and Ka band receivers covering the spectral range between 8 and 40 GHz, with angular resolution of ranging from 0.3′ to 1′.5 FWHM, with polarization capability and very competitive sensitivity. The GBT has already undertaken observational projects related to AME. For instance, Mason et al. [9] placed tight upper limits on the polarization of the anomalous source L1622 (previously observed by Finkbeiner et al. [34] with the 140-foot GBT) and Murphy et al. [25] used the GBT for the first extragalactic evidence of AME. The NGC 6946 result was obtained using the Caltech Continuum Backend (CCB) on the (GBT). Observations were obtained demodulating the beam-switched signal to remove atmospheric fluctuation and receiver gain variations [35]. Average beam in the 26–40 GHz with the GBT is 25″. Observational pressure on the GBT is strong, so not many large programs are usually accepted [36]. However, single source of known or unknown AME usually fits the observational management of the GBT.

*3.2. Effelsberg.* The 100 m Effelsberg telescope is the largest European telescope. No active surface is implemented on the primary mirror, while eight actuators are installed on the secondary mirror. Its nominal observational frequencies are between 408 MHz and 86 GHz thanks to the special compensating support structure used. Its fleet covers the frequency of interest for AME measurements, but continuum observations are possible only at 1.3–1.7 GHz, 2.6 GHz, 4.85 GHz, 8.35 GHz, 10.5 GHz, and 32 GHz. To our knowledge no observational program has been undertaken in the field of AME. Due to its location, radio frequency interference may be an issue. The very efficient beam-switching receivers may however be particularly adequate for high-resolution observations of AME regions.

*3.3. Parkes.* One very scientifically successful single-dish telescope is the Parkes 64 m telescope. Among the reasons of its success are the availability of a state-of-the-art L-band 13-beam receiver, digital backends, and its particular radio quiet and geographical location. Parkes is equipped with a consistent receivers fleet, from the MARS to the 13 mm receiver, which allows to cover the spectral range of interest for the AME. Parkes has recently undergone a program for observing AME sources (Cruciani et al. in preparation) [32, 37] and is particularly adequate for such a program.

*3.4. Sardinia Radio Telescope.* The new upcoming giant single-dish steerable radio telescope is the Sardinia Radio Telescope (SRT). SRT is a new 64 m state-of-the-art radio-telescope. The most important feature of SRT is the active surface which enables the SRT to observe the sky up to 100 GHz. It is located in Sardinia in a particularly radio quite zone. One of the special characteristics of the SRT is the availability of three different focal positions allowing to efficiently manage different kind of receivers. The spectral rage of interest for AME observation is covered by several Ku, K, and Ka receivers with $2 \times 2000$ MHz bandwidth each, and mjy$\sqrt{s}$-level sensitivity, all positionable in the Gregorian focus. The modern design and used technology make the SRT among the most competitive Giant Radio Telescope ever built. Different observational techniques can be implemented like position switching, beam switching, frequency switching, and the secondary mirror wobbling thanks to the 6 electromechanic actuators enabling fast tilting. Mapping techniques scanning and on the fly are also possible. Continuum observations are planned with the multibeam receiver in the K band. 7 beams in the 18–26 GHz range will enable the observer to track the AME where its spectrum shows considerable different features for the other well-known dominant emission mechanisms in the microwave band like free-free or synchrotron.

In order to estimate the observing time to achieve maps of microwave emitting regions with the different telescopes, we have estimated the nominal observing time required to map a $30' \times 30'$ region with *rms* noise down to the mJy/beam level, enough for typical anomalous emitting regions of several mJy/arc min$^2$. Assuming 1-minute long scans to keep a typical 30 s overhead between consecutive scans within an acceptable fraction, and assuming 1/3 FWHM spaced scans and one full coverage along each of two orthogonal directions (RA and DEC) to make basket-weaving map-making effective, we find that, for the 13 GHz measurements, 1 full scan of 2.5 hours allows to reach the 1.8 mJy/beam level for the Parkes telescope, 1 full scan of 5 hours allows to reach the 0.6 mJy/beam level for the GBT, and 1 full scan of 3 hours allows to reach the 0.35 mJy/beam level for the SRT. As for the 23 GHz measurements, we find that, with similar scanning procedures, we have a 4-hour scan allows to reach the 2 mJy/beam level for the Parkes telescope, a 3-hour scan would allow to reach the 0.3 mJy/beam level for the GBT, and a 5-hour scan allows to reach the 1.5 mJy/beam for the SRT. Different scan lengths reflect the difference in the observing beams and the necessity to preform different number of single scans in order to uniformly cover the observed region. We stress, however, that the real integration time and the final flux *rms* level depend also on several other factors such as system instabilities, $1/f$ noise, and details of map-making. The specified times, while certainly promising for these observations, should thus be interpreted as optimistic.

## 4. Conclusions

We have investigated the role that high angular resolution measurements of the AME will have in the near future. Comparing the instantaneous sensitivity among the cited four instruments [36] shows that all of them would be incredibly useful for AME measurements. In particular, the GBT obviously shows the best performance in terms of angular resolution thanks to its size and the SRT is competitive with all the other instrument often overtaking them all. Several projects to the mentioned instruments are adviceable in order to unveil the AME mechanisms.

## Acknowledgments

The authors acknowledge the use of the MIPS 24-micron data from the NASA/IPAC Infrared Science Archive. They thank Ricardo Génova-Santos for comments and suggestions. We thank the anonymous referee for useful comments that improved the paper.